\documentclass[12pt,a4paper]{article}
\usepackage[a4paper, margin=2.5cm]{geometry}
\usepackage[T1]{fontenc}
\usepackage[utf8]{inputenc}
\usepackage{amsmath,amssymb,amsthm}
\usepackage[numbers,sort&compress]{natbib}
\usepackage{hyperref}
\usepackage{booktabs}   
\usepackage{graphicx}   
\usepackage{subcaption} 
\usepackage{a4}
\newcommand{\Schwartz}{\mathcal{S}}   
\newcommand{\Four}{\mathcal{F}}       
\newcommand{\Rad}{\mathcal{R}}        
\newcommand{\Hil}{\mathcal{H}}        

\newcommand{\abs}[1]{\lvert#1\rvert}

\DeclareMathOperator{\sgn}{sgn}

\theoremstyle{plain}

\newtheorem{proposition}{Proposition}

\newtheorem{lemma}{Lemma}

\theoremstyle{definition}
\newtheorem{definition}{Definition}
\newtheorem{properties}{Properties}

\begin{document}

\title{A Radon reconstruction with data from an enlarged region of interest}
 
\author{J\"{o}ran Bergh%
    \thanks{Department of Mathematical Sciences,
      (Chalmers University of Technology and) University of Gothenburg,
      SE-412\,96 G\"{o}teborg, Sweden;
       \texttt{bergh@chalmers.se or joran.bergh@gu.se}.}
    \and
    Martin Lindberg%
    \thanks{Independent researcher, \texttt{martin.l.lindberg@icloud.com};
      Hantverkarev\"{a}gen 6,
      SE-237\,36 Bj\"{a}rred, Sweden.}}
\date{}
\maketitle

\begin{abstract}
In x-ray tomography it is desirable to minimize radiation exposure by
restricting measurements to a region of interest and its immediate
surroundings.
We propose an approximate reconstruction based on a convolution formula that
reconstructs mean values of the unknown function from truncated Radon
data.
The key ingredient is a reconstruction kernel derived from the normalized
characteristic function of a disk.
We derive explicit error estimates for this cut-off approximation and show
that, for a compactly supported function, the cut-off essentially only causes
the addition of a constant in the (not too small) region of interest.
A numerical comparison against standard filtered back-projection,
implemented in \textsc{Matlab}, demonstrates that our method, using 'estimated overall mean,'
achieves competitive accuracy while restricting to data from the enlarged region of interest only.
\end{abstract}


\section{Introduction}\label{sec:intro}

The aim of this note is to describe an approximate Radon reconstruction
technique from data in an enlarged region of interest only.
Our main tool is a mean value formula.
There is a medical background: in x-ray computed tomography,
it is desirable to minimize the radiation dose to the patient by restricting
measurements to the lines intersecting a region of interest and a prescribed
neighborhood of it.

The standard Radon inversion, or back-projection, formula in the plane,
\begin{equation}
  f(x) = 1/(4\pi) \int_{\mathbf{S}^1}
    D\Hil\Rad_\theta f \,(x\cdot\theta) \,d\theta, \label{eq:backproj}
\end{equation}
involves the Hilbert transform $\Hil$, which is non-local, and so obstructs
local reconstruction from Radon data passing through a region of interest only.
Consequently, computing $f(x)$ for all $x$ in the region of interest
in principle requires Radon data from all lines, not only those intersecting
the region of interest.
This non-locality underlies the Interior Problem: Radon data from
lines intersecting a disk are in general insufficient for unique
reconstruction of $f$ inside that disk~\cite{natterer1986}.
The Support Theorem~\cite{helgason1999} shows, conversely, that if the
Radon data vanish outside a convex region, the function vanishes there as
well. 

Several approaches to region of interest reconstruction in computed tomography
have been developed, referred to as local tomography.
These are often based on the convolution-backprojection duality: 
\begin{equation}
    f * R^* g = R^* (Rf * g),
\end{equation}
where $R^*$ is the back-projection operator.
The idea is to choose $g$ so that $R^* g$ is an approximation of the identity.

Our starting point will be such a convolution formula,
which reconstructs mean values of $f$:
\begin{equation}\label{eq:conv}
  f * \varphi \,(x) = 1/(4\pi) \int_{\mathbf{S}^1}
    \Rad_\theta f * D\Hil\Rad_\theta\varphi\,(x\cdot\theta) \, d\theta,
\end{equation}
where $\varphi$ is a specific mean value density.
We recall the mathematics behind this formula in Section~\ref{sec:math}.
We note that, in this case,
the sequence $\varphi_n = n^2\varphi(n \cdot)$ 
tends to the Dirac $\delta$, and
the values  $f*\varphi_n (x)$ tend to  $f(x)$. Thus this
convolution formula contains the standard back-projection as a limiting case.
To our knowledge, formula~\eqref{eq:conv} has not previously been used in
the enlarged region of interest reconstruction context.

We will use the normalized characteristic function of a disk as
our mean value density.
This density has two key advantages: its Radon transform is independent of
$\theta$ and is given in closed form, and the resulting reconstruction
kernel $D\Hil\Rad_\theta\varphi$ has a simple closed form and decays,
outside the interval $[-1, 1]$, as the inverted square of the distance
to the disk.
Damping rates are our main backdrop; they are treated in
Section~\ref{sec:damping}.

The main contributions of this paper are as follows.
\begin{enumerate}
  \item We extend formula~\eqref{eq:conv} to the density
        $\varphi = \pi^{-1}\mathbf{1}_D$ (Proposition~2), which requires
        care since $D\Hil\Rad_\theta\varphi$ is not square integrable.
  \item We prove that the reconstruction kernels  $D^k\Hil\Rad_\theta\varphi(s)$
        decay exactly as $|s|^{-(k+1)}$
        for any mean value density $\varphi \in \Schwartz$ 
        (Proposition~3), and verify the $|s|^{-2}$ rate for
        our chosen $\varphi$.
  \item We derive cut-off error bounds showing that the truncation error is
        bounded by $C_\varphi\,d^{-2}\,\int_{\bf S^1}\int_{|s|>d}|\Rad_\theta f(s)| ds\, d\theta$
        (Proposition~6), and that the cut-off error function $f_o$ is
        analytic in the interior of the region of interest (Proposition~7).
  \item We introduce two mean-value corrections for the cut-off
        error --- an \textit{estimated overall mean} and \textit{local means} ---
        that reduce it by subtracting and restoring a determined constant
        (global or per-pixel, respectively). We further indicate a weighting to
        mitigate the artifacts at the cut-off boundary.
  \item We compare the reconstructions against standard filtered back projection in \textsc{Matlab}
        using the Modified Shepp--Logan Phantom, demonstrating competitive accuracy
        from enlarged-region-of-interest data only.        
\end{enumerate}

For discretization and computation, we use the parallel scanning geometry
with measurement lines at the same spacing as the pixel lattice spacing.
We interpret the mean value density $\varphi$ as centered at each lattice
point, with its radius equal $1/\sqrt{\pi}$ (unit area with the pixel distance 1).
However, in the theoretical sections we work with radius $1$ for simplicity.
The reconstructed values are the computed mean values around the lattice points.

A key aspect of the method is the damping factor: the contribution from
Radon data at distance $p$ lattice points away decays as $1/p^2$ or faster
(Proposition~\ref{prop:damping}). This allows truncation of the data to an
enlarged region of interest with a manageable error. The damping persists
as the mean value density approaches the Dirac delta, so the method interpolates
smoothly between reconstruction of means and pointwise inversion.

The paper is organized as follows. In
Section~\ref{sec:math}, we recall the mathematical background: notation, the convolution
formula, and note its stability for perturbations in the Radon data. 
Then we specialize to the planar disk: derive the closed-form
kernel, prove the convolution formula for our chosen density,
establish the  kernel damping rates, and treat the
higher-dimensional analogues. In
Section~\ref{sec:approximations}, we discuss the two approximations inherent in the method ---
mean-value substitution and data cut-off --- and introduce the two
procedures using means. Finally, in
Section~\ref{sec:numerics}, we present the numerical experiments.

\subsubsection*{Related work}

Convolution methods are extensively used in Radon reconstruction contexts;
we refer to~\cite{natterer1986} for a comprehensive account of reconstruction issues.
Specifically, the use of the characteristic function of a disk in~\eqref{eq:conv} has been
documented in~\cite{nievergelt1986}

Truncated filtered back-projection simply cuts off the data at the
region of interest boundary and accepts the resulting artefacts~\cite{natterer1986}.
In Section~\ref{sec:roi} we return to an empirical observation made in~\cite{natterer1986}
in this context.

Another approach~\cite{defrise2006}, uses a truncated Hilbert transform along detector
rows together with a priori knowledge (such as a known subregion where $f$
vanishes) to achieve exact interior reconstruction.
Further developments appear in~\cite{kudo2008,ye2007}.

There is also an application of 'compressed sensing' algorithms to Fourier backprojection methods. See~\cite{candes2006}
and~\cite{donoho2006} for basic results. 
The reconstruction matrix of the present mean value method
violates the 'close-to-orthogonality' criterion there. 

We have found no record of~\eqref{eq:conv} being used in the context of
region-of-interest reconstruction.

\section{Mathematical background}\label{sec:math}

First some notation.
We denote by $\Schwartz$ the usual Schwartz class of rapidly decreasing
smooth functions on $\mathbf{R}^n$: a function and any of its derivatives
decrease faster than the inverse of any polynomial at infinity.
For simplicity we assume throughout that functions belong to $\Schwartz$,
unless stated otherwise; most statements hold under much weaker assumptions.
We write $x\cdot y := x_1 y_1 + \cdots + x_n y_n$ for the inner product
in $\mathbf{R}^n$.

\begin{definition}
For $f \in \Schwartz$ the Fourier transform is defined by
\[
  \Four f(\xi) := \int_{\mathbf{R}^n} e^{-2\pi i \xi\cdot x} f(x) \, dx.
\]
\end{definition}

\begin{definition}
For $f \in \Schwartz$ the Radon transform
$\Rad\colon\Schwartz(\mathbf{R}^n) \to \Schwartz(Z)$
is defined, for $\theta \in \mathbf{S}^{n-1}$ and $s \in \mathbf{R}$, by
\[
  \Rad f(\theta,s) := \int_{x\cdot\theta = s} f(x) \, dx,
\]
where $Z := \mathbf{S}^{n-1} \times \mathbf{R}$ is the unit cylinder,
equipped with the product measure $d\theta\,ds$.

\end{definition}
We write $\Rad_\theta f(s)$ for $\Rad f(\theta,s)$ to denote the
Radon transform of $f$ with respect to $s$ for fixed $\theta$.
The following symmetry holds: $\Rad_{-\theta} f(-s) = \Rad_\theta f(s)$.
We note the relation $\Rad_\theta D_i f = \theta_i D \Rad_\theta f$,
where $D_i$ is the $i$-th partial derivative and $D$ denotes the derivative
with respect to $s$; this follows directly from the definition.
With $\nabla f$ denoting the gradient of $f$ we also note that
 for $x \in \mathbf{R}^n$:
\[
  \Rad_\theta(x \cdot \nabla f) = (x\cdot\theta) \, D\Rad_\theta f.
\]

\begin{definition}
The \textit{back-projection operator} $\Rad^*\colon\Schwartz(Z)
\to \Schwartz(\mathbf{R}^n)$ is defined by
\[
  \Rad^* g\,(x) := \int_{\mathbf{S}^{n-1}} g(\theta,\, x\cdot\theta)
    \, d\theta.
\]
\end{definition}
The operators $\Rad$ and $\Rad^*$ are mutually adjoint on the
corresponding $L^2$ spaces.

\begin{definition}
For $f \in \Schwartz$ the Hilbert transform is given by
\[
  \Hil f(s) := \frac{1}{\pi} \int_{\mathbf{R}} \frac{f(t)}{s-t} \, dt.
\]
\end{definition}

Recall that for $f, g \in \Schwartz$:
\begin{itemize}
  \item $\Four\Hil f = -i\operatorname{sgn}(\cdot)\,\Four f$,
        where $\operatorname{sgn}(\sigma)$ denotes the sign of $\sigma$;
  \item $\Hil^{2m+1} f = (-1)^m \Hil f$;
  \item $D\Hil f = \Hil D f$;
  \item $\Hil(f * g) = (\Hil f)*g = f*(\Hil g)$;
  \item $D(f * g) = (Df)*g = f*(Dg)$.
\end{itemize}

\begin{lemma}[Fourier slice equality]\label{lem:fourier-slice}
Let $f \in \Schwartz$.  Then for
$\sigma \in \mathbf{R}$, $\theta \in \mathbf{S}^{n-1}$:
\[
  \Four\Rad_\theta f(\sigma) = \Four f(\sigma\theta).
\]
\end{lemma}
This well-known equality is obtained by a change in the order of integration.

The following convolution formula appears in equivalent forms
in~\cite{helgason1999,natterer1986}, and constitutes the basis of our
reconstruction method.
The standard Radon inversion formula corresponds to the limiting case
$g = \delta$, and a proof is obtained in the same way as that for the convolution formula.

\begin{proposition}[Convolution formula]\label{prop:convolution}
Let $f$ and $g$ be functions in $\Schwartz$.  Then:
\begin{equation}\label{eq:conv-prop}
  f * g \,(x) = 1/2 \int_{\mathbf{S}^{n-1}}
    \Rad_\theta f * \left(1/ (2\pi) \Hil D  \right)^{n-1}
    \Rad_\theta g\,(x\cdot\theta) \, d\theta.
\end{equation}
\end{proposition}

\begin{proof}
Using polar coordinates and Lemma~\ref{lem:fourier-slice}, we get:
\begin{align*}
  f * g \,(x)
    &= \int_{\mathbf{R}^n} e^{2\pi i x\cdot\xi}\,
       \Four f(\xi)\, \Four g(\xi) \, d\xi \\
    &= 1/2 \int_{\mathbf{S}^{n-1}} \int_{\mathbf{R}}
       e^{2\pi i \sigma x\cdot\theta}\,
       \Four f(\sigma\theta)\,
       \bigl(\sigma\operatorname{sgn}(\sigma)\bigr)^{n-1}\,
       \Four g(\sigma\theta) \, d\sigma \, d\theta \\
    &= 1/2 \int_{\mathbf{S}^{n-1}}
       \Rad_\theta f * \left(1/(2\pi)\Hil D \right)^{n-1}
       \Rad_\theta g\,(x\cdot\theta) \, d\theta.
\end{align*}
\end{proof}

We remark that the convolution formula is stable for perturbations in $ \Rad_\theta f$,
due to, {\it e.g.}, the inequality:
\begin{align*} \big| 1/2 \int_{\mathbf{S}^{n-1}}
  &
       \Rad_\theta f * \left(1/(2\pi)\Hil D \right)^{n-1}
  \Rad_\theta g\,(x\cdot\theta) \, d\theta\, \big| \\
  &\leq
       1/2 \sup_{\theta,\, s}|\Rad_\theta f(s)|
       \int_{\mathbf{S}^{n-1}}\int_{\mathbf R}
       \big| \left(1/(2\pi)\Hil D \right)^{n-1}\Rad_\theta g(s)|\, ds\, d\theta\, \big| \, .
\end{align*}

In dimension $n = 2$, with $g = \varphi$, the convolution formula specializes to
\[
  f * \varphi \,(x) = \frac{1}{4\pi} \int_{\mathbf{S}^1}
    \Rad_\theta f * D\Hil\Rad_\theta g\,(x\cdot\theta) \, d\theta,
\]
which is formula~\eqref{eq:conv} in the Introduction for $\varphi\in\Schwartz$.

However, our results on cut-off and reconstruction, in Section~\ref{sec:approximations},
hold for general mean value densities: non-negative functions with integral equal to~$1$.
Using a standard regularization, our chosen density may be approximated by
a function in $\Schwartz$.
We made our specific choice of density in order to obtain relative
computational ease: all functions involved appear in closed form.

\subsection{Mean values over disks}\label{sec:disk}

We now treat the case $n = 2$.
Our mean value density $\varphi$ takes the value $1/\pi$ on the unit disk
$D$ and the value $0$ elsewhere.
(The symbol $D$ denotes both the unit disk and the derivative operator;
the meaning will be clear from context.)
We write $\varphi = \pi^{-1}\mathbf{1}_D$ and record the following.

\begin{properties}[mean value over disk]\label{prop:disk}
For $\varphi = \pi^{-1}\mathbf{1}_D$ we record:
\begin{align}
  \Rad_\theta\varphi(s) &= \,
    2/\pi \left(1-s^2\right)^{1/2}\,\mathbf{1}_{[-1,1]}(s) \label{eq:rad-phi} \\
  D\Hil\Rad_\theta\varphi(s) &=
  \begin{cases}
    2/\pi                               & \text{if } |s| < 1 \\
    2/\pi\left(1-|s|(s^2-1)^{-1/2}\right)  & \text{if } |s| > 1
  \end{cases} \label{eq:der-hil-rad-phi}
\end{align}
Further, we may rewrite: \[ 2/\pi\left(1-|s|(s^2-1)^{-1/2}\right) =
\frac{2}{\pi(|s|+(s^2-1)^{1/2})(s^2-1)^{1/2}}  \, . \]
\end{properties}

Property~\eqref{eq:rad-phi} is immediate from the definition of the Radon transform.
We further have the following intermediate property:
\begin{equation}\label{eq:hil-rad-phi}
  \Hil\Rad_\theta\varphi(s) =
  \begin{cases}
    2s/\pi                                    & \text{if } |s| < 1 \\
    2/\pi\left(s-\sgn(s)(s^2-1)^{1/2}\right)  & \text{if } |s| > 1
  \end{cases}
\end{equation}
which we use for quadrature in the numerical experiments in Section~\ref{sec:numerics}.

For the inner part of property~\eqref{eq:hil-rad-phi} ($\abs{s} < 1$),
we note that the classical Chebyshev polynomials $T_l$ of the first kind
and $U_l$ of the second kind satisfy
\[
  -1/\pi \int_{-1}^1
    \frac{T_l(t)(1-t^2)^{-1/2}}{s-t} \, dt
  = U_{l-1}(s), \qquad |s| < 1,
\]
in which we take $l=1$: $U_0 = 1$, $T_1(t) = t$.
(This equality translates, by a change of variable, to a formula for the
conjugate function on the circle $\mathbf{S}^1$; for $l=1$ it expresses
that the conjugate of $\cos x$ is $\sin x$.)

For the outer part of property~\eqref{eq:hil-rad-phi} ($\abs{s} > 1$),
we use the known relation
\[
  \frac{1}{\pi}\int_{-1}^1 \frac{(1-t^2)^{1/2}}{s-t} \, dt
  = s - \operatorname{sgn}(s)(s^2-1)^{1/2},
  \qquad |s| > 1,
\]
established by the substitution $u = \left((1+t)/(1-t)\right)^{1/2}$.

\begin{figure}[h]
  \centering
  \includegraphics[width=0.8\textwidth]{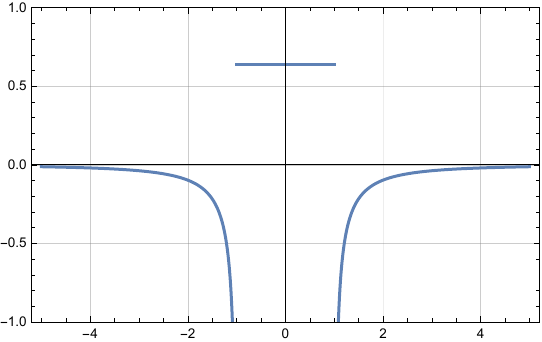}
  \caption{The reconstruction kernel $D\Hil\Rad_\theta\varphi(s)$}
  \label{fig:rec-kern}
\end{figure}

We will refer to  the function $D\Hil\Rad_\theta\varphi$ as the
\textit{reconstruction kernel}, see Figure~\ref{fig:rec-kern}, and observe that it
is not locally square integrable, which blocks the use of the direct
Fourier method (employed above to prove~\eqref{eq:conv-prop}) with the operator
$D\Hil$ falling on $\Rad_\theta\varphi$.
From~\eqref{eq:der-hil-rad-phi}, this function has two singularities of order
$(\cdot)^{-1/2}$ at $\pm 1$, so it is integrable but not square
integrable.

We now extend the formula to our density $\varphi$.
\begin{proposition}[Extension to the disk density]\label{prop:extension}
For $f \in \Schwartz$ and $\varphi = \pi^{-1}\mathbf{1}_D$:
\begin{equation}\label{eq:conv-disk}
  f * \varphi \,(x) = \frac{1}{4\pi} \int_{\mathbf{S}^1}
    \Rad_\theta f * D\Hil\Rad_\theta\varphi\,(x\cdot\theta) \, d\theta.
\end{equation}
\end{proposition}

\begin{proof}
Since convolutions are translation invariant, it suffices to prove the
identity at $x = 0$. Note that $D\Hil\Rad_\theta\varphi(s)$ is even in $s$. We will thus show
that:
\[
  \int_{\mathbf{R}^2} f(x)\varphi(x) \, dx
  = \frac{1}{4\pi} \int_{\mathbf{S}^1} \int_{\mathbf{R}}
      \Rad_\theta f(s)\, D\Hil\Rad_\theta\varphi(s) \, ds \, d\theta.
\]
Since $D\Hil\Rad_\theta\varphi$ is not square integrable, the Fourier argument from
Proposition~\ref{prop:convolution} does not apply directly.
Instead, we use an integration by parts.

Since $\Rad_\theta\varphi$
and $\Hil\Rad_\theta\varphi$ are both square integrable,
the left hand side can be rewritten, using the Fourier argument to prove 
Proposition~\ref{prop:convolution} :
\[
  \int_{\mathbf{R}^2} f(x)\varphi(x) \, dx
  = -\frac{1}{4\pi} \int_{\mathbf{S}^1} \int_{\mathbf{R}}
      D\Rad_\theta f(s)\, \Hil\Rad_\theta\varphi(s) \, ds \, d\theta.
\]
Now integrate by parts the factor  $D\Hil\Rad_\theta\varphi$:
the derivative $D$ transfers from $\Rad_\theta f$ to $\Hil\Rad_\theta\varphi$, giving
(boundary terms vanish):
\[
  \int_{\mathbf{R}} D\Rad_\theta f(s)\, \Hil\Rad_\theta\varphi(s) \, ds
  = - \int_{\mathbf{R}} \Rad_\theta f(s)\, D\Hil\Rad_\theta\varphi(s) \, ds.
\]

This is justified because $D\Hil\Rad_\theta\varphi$ is integrable:
by~\eqref{eq:der-hil-rad-phi} its singularities at $\pm 1$ are of the order
$(\cdot)^{-1/2}$, which implies integrability.
The result follows.
\end{proof}

\subsection{Damping}\label{sec:damping}

Decay rates of reconstruction kernels are our focus here.

\begin{proposition}[Damping]\label{prop:damping}
Assume $\varphi \in \Schwartz$ with $\int \varphi = 1$.
Then for $k \geq 0$:
\[
  \lim_{|s|\to\infty} |s|^{k+1}
  \bigl|\Hil D^k \Rad_\theta\varphi(s)\bigr| = \frac{k!}{\pi}.
\]
For our chosen density $\varphi = \pi^{-1}\mathbf{1}_D$:
\[
  \lim_{|s|\to\infty} s^2
  \bigl|\Hil D\Rad_\theta\varphi(s)\bigr| = \frac{1}{\pi}.
\]
\end{proposition}

\begin{proof}
Let $s \neq 0$ and $\psi \in \Schwartz$.  The identity
\[
  s \int \frac{\psi(t)}{s-t} \, dt
  = \int \psi(t) \, dt + \int \frac{t\psi(t)}{s-t} \, dt
\]
may be iterated to give, for $k \geq 0$:
\begin{align*}
  s^{k+1} \int \frac{\psi(t)}{s-t} \, dt
    &= s^k \int \psi(t) \, dt + \cdots
       + \int t^k\psi(t) \, dt
       + \int \frac{t^{k+1}\psi(t)}{s-t} \, dt.
\end{align*}
If the first $k$ moments of $\psi$ vanish, but
$\int t^k\psi(t) \, dt \neq 0$, the Hilbert transform decays exactly as
$|s|^{-(k+1)}$ and
\[
  \lim_{|s|\to\infty} |s|^{k+1} |\Hil\psi(s)|
  = \frac{1}{\pi}\left|\int t^k\psi(t) \, dt\right| \neq 0.
\]
(The last integral tends to $0$ as $|s| \to \infty$, which is immediate
when $\psi$ has compact support, and follows in general from a Fourier
representation.)
Setting $\psi = D^k\Rad_\theta\varphi$ with $\int\varphi = 1$, the first
$k$ moments of $\psi$ vanish by partial integration ($0 \leq l \leq k;
\ \int \Rad_\theta\varphi(t) \, dt = \int \varphi(x) \, dx$):
\begin{align*}
  \int t^l D^k&\Rad_\theta\varphi(t) \, dt = ... \\
    &= \max(l-k+1,\,0)\,(-1)^k k! \int \varphi(x) \, dx \\
    &= \max(l-k+1,\,0)\,(-1)^k k!,
\end{align*}
which proves the first statement.
The second follows directly from~\eqref{eq:der-hil-rad-phi}.
\end{proof}

\subsection{Mean values over balls}\label{sec:balls}

In dimension $n > 2$, analogues of
\eqref{eq:rad-phi}--\eqref{eq:der-hil-rad-phi} and
Proposition~\ref{prop:extension} hold, substituting $B^n$ for $D = B^2$.
We only sketch the proofs, since our present focus is $n = 2$.

For our mean value density
$\varphi = |B^n|^{-1}\mathbf{1}_{B^n}$,
we have ($1 \leq l \leq n-1$):

\begin{properties}\label{prop:ball}
\begin{align}
  \Rad_\theta\varphi(s)
    &= \frac{|B^{n-1}|}{|B^n|}(1-s^2)^{(n-1)/2}\,\mathbf{1}_{B^1}(s),
    \label{eq:rad-ball} \\
  D^l\Hil\Rad_\theta\varphi(s)
    &= p_{n-1-l}(s), \quad |s| < 1,
    \label{eq:dhR-ball-inner} \\
  D^l\Hil\Rad_\theta\varphi(s)
    &= c_{n,l}\int_{-1}^1
       \frac{(1-t^2)^{(n-1)/2}}{(s-t)^{l+1}} \, dt,
     \quad |s| > 1,
    \label{eq:dhR-ball-outer}
\end{align}
\end{properties}
where $c_{n,l} = \pm\pi^{-1}\,l!\,|B^{n-1}|/|B^n|$
and $p_{n-1-l}$ is a polynomial of degree $n-1-l$.

Clearly, \eqref{eq:rad-ball} follows directly from the definition.
When $n = 2k+1$ is odd, there is no operative Hilbert transform
in~\eqref{eq:conv-prop}, so \eqref{eq:dhR-ball-inner}
and~\eqref{eq:dhR-ball-outer} are not relevant; it follows
from~\eqref{eq:rad-ball} that $D^{2k}\Rad_\theta\varphi(s)$ is piecewise
constant.

We now consider only even dimensions $n = 2k$, when
\eqref{eq:dhR-ball-inner} and~\eqref{eq:dhR-ball-outer} come into play.
Formula~\eqref{eq:dhR-ball-outer} may be obtained in closed form by the
substitution $u = \bigl((1+t)/(1-t)\bigr)^{1/2}$.
The decay in~\eqref{eq:dhR-ball-outer} for large $|s| > 1$ is of order
$|s|^{-(l+1)}$.
The left-hand side of~\eqref{eq:dhR-ball-inner}, omitting constant factors,
is
\[
  D^l \int_{-1}^1 \frac{(1-t^2)^{n/2-1/2}}{s-t} \, dt.
\]
Since any polynomial is a linear combination of Chebyshev polynomials $T_l$
up to the same degree, and using the Chebyshev formula cited in
Section~2, this expression is a polynomial of degree $n-1-l$ for
$|s|<1$.

From~\eqref{eq:dhR-ball-outer} with $n = 2k$,
\[
  \int_{-1}^1 \frac{(1-t^2)^{k-1/2}}{(s-t)^{l+1}} \, dt
\]
has singularities at $\pm 1$ when $l \geq k$.
For $l = k$ the integral is integrable but not for $l > k$; accordingly,
we must let $D^{k-1}$ derivatives fall on $\Rad_\theta f$ in the
reconstruction formula.
The decay for $l = k$ is of order $|s|^{-(k+1)}$ for large $|s|$.
The next lemma describes the behaviour at the singular points $s = \pm 1$.

\begin{lemma}\label{lem:singularity}
For our density $\varphi$ with $k \leq l \leq 2k-1 = n-1$ and $|s| > 1$:
\[
  0 < A \leq
  \frac{\bigl|D^l\Hil\Rad_\theta\varphi(s)\bigr|}{(|s|-1)^{k-l-1/2}}
  \leq B.
\]
\end{lemma}

\begin{proof}
It suffices to consider $s = 1$, since $D^l\Hil\Rad_\theta\varphi$ is
even when $l$ is odd and odd when $l$ is even.  For $s > 1$:
\[
  \bigl|D^l\Hil\Rad_\theta\varphi(s)\bigr|
  = C \int_{-1}^1 \frac{(1-t^2)^{k-1/2}}{(s-t)^{l+1}} \, dt.
\]
The substitution $1-t = u(s-1)$ gives
\[
  \frac{C}{(s-1)^{k-l-1/2}}
  \int_0^{2/(s-1)}
  \frac{u^{k-1/2}(2-u(s-1))^{k-1/2}}{(1+u)^{l+1}} \, du,
\]
from which the bounds $A$ and $B$ follow.
\end{proof}

We have sketched the proof of the following generalization of
Proposition~\ref{prop:extension}.

\begin{proposition}\label{prop:ndim}
For $n = 2k+1$ and $f \in \Schwartz$:
\[
  f * \varphi(x) = \frac{1}{2} \int_{\mathbf{S}^{2k}}
    \Rad_\theta f *
    (-1)^k\!\left(\frac{D}{2\pi}\right)^{\!2k}
    \Rad_\theta\varphi\,(x\cdot\theta) \, d\theta,
\]
where $(-1)^k(D/2\pi)^{2k}\Rad_\theta\varphi = p_0\,\mathbf{1}_{B^1}$.
For $n = 2k$ and $f \in \Schwartz$:
\[
  f * \varphi(x) = \frac{(-1)^{k-1}}{2} \int_{\mathbf{S}^{2k-1}}
    \left(\frac{D}{2\pi}\right)^{k-1}
    \Rad_\theta f *
    \left(\frac{D}{2\pi}\right)^{k} \Hil\Rad_\theta\varphi\,(x\cdot\theta) \, d\theta.
\]
\end{proposition}

\section{Approximations and a region of interest}\label{sec:approximations}

Let us first note that a Radon reconstruction from sampled data implicitly
assumes that the function being reconstructed has negligible content at
scales smaller than twice the distance between adjacent lines.
By the Fourier slice equality (Lemma~\ref{lem:fourier-slice}), this means
$\Four f(\xi)$ is negligible for $|\xi| > 1/2$ (in this distance unit);
otherwise alias effects distort the reconstruction.
Recall further that, by the Paley--Wiener theorem, a function and its
Fourier transform cannot both have compact support.

We consider here the two distinct approximations that appear in the present
reconstruction method.
The first is common: substituting the mean value of $f$ over a small disk
around a point for the function value at that point.
The second is cutting off Radon data when calculating mean values.
We prove a rough estimate of the cut-off error, and we discuss ways to
reduce this error in reconstructions from data in an enlarged region of
interest only.

For simplicity, we take $f \in \Schwartz$ throughout this section;
our statements hold under much weaker assumptions, and are easily seen to
hold for general mean value densities.

\subsection{Mean value and small scales}

Substituting a mean value for a point value, the error depends to first
order on the gradient of $f$ in the averaging domain.

\begin{proposition}\label{prop:meanvalue}
For $f \in \Schwartz$ and any $a \in \mathbf{R}^n$, $r > 0$:
\[
  \left|\frac{1}{r^n|B^n|} \int_{|x-a|<r} f(x) \, dx - f(a)\right|
  \leq r \sup_{|x-a|<r} |\nabla f(x)|.
\]
\end{proposition}

\begin{proof}
\begin{align*}
  &\left|\frac{1}{r^n|B^n|} \int_{|x-a|<r} f(x)\,dx - f(a)\right| \\
  &\qquad\leq \frac{1}{r^n|B^n|} \int_{|x-a|<r} |f(x)-f(a)| \, dx \\
  &\qquad= \frac{1}{r^n|B^n|} \int_{|x-a|<r}
     \left|\int_0^1 \nabla f(a+t(x-a))\cdot(x-a) \, dt\right| dx \\
  &\qquad\leq r \sup_{|x-a|<r} |\nabla f(x)|. \qquad
\end{align*}
\end{proof}

In addition, if we assume $\Four f(\xi) = 0$ for $|\xi| > 1/2$, as
discussed above, we have over-all estimates.
For perspective, here are two such estimates:
\begin{align*}
  |f(x)|
    &\leq \int_{|\xi|<1/2} |\Four f(\xi)| \, d\xi
     \leq C \int_{\mathbf{R}^n} |f(y)| \, dy, \\
  |\nabla f(x)|
    &\leq C
     \int_{\mathbf{R}^n} |f(y)| \, dy\, .
\end{align*}
For $n=2$, both constants are of the magnitude order $1$.

\subsection{Cut-off error}\label{sec:cutoff}

We restrict here to dimension $n = 2$ (with a remark on $n > 2$ at the end)
and our chosen mean value density $\varphi$, unless stated otherwise.
We let $d$ denote the cut-off distance from the centre of the unit disk
$D$, placed at the origin.
The translation to a general centre is obvious.

We estimate the error in the mean value $f*\varphi(0)$ introduced by
cutting off data from lines at distance greater than $d > 2$.
For $x=0$ and $n=2$, since $D\Hil\Rad_\theta\varphi(s)$ is even in $s$,
formula~\eqref{eq:conv} reduces to:
\[
  f * \varphi(0) = \frac{1}{4\pi} \int_{\mathbf{S}^1} \int_{\mathbf{R}}
    \Rad_\theta f(s)\, D\Hil\Rad_\theta\varphi(s) \, ds \, d\theta.
\]

\begin{proposition}[Cut-off error bound]\label{prop:cutoff}
For $f \in \Schwartz$, the error in $f*\varphi(0)$ from cutting off data
at distance $d > 2$ satisfies:
\begin{align}
  &\left|\frac{1}{4\pi}\int_{\mathbf{S}^1}\int_{|s|>d}
    \Rad_\theta f(s)\, D\Hil\Rad_\theta\varphi(s)
    \, ds\, d\theta\right|
  \leq \frac{C_\varphi}{4\pi d^2}
    \int_{\mathbf{S}^1}\int_{|s|>d} |\Rad_\theta f(s)| \, ds\, d\theta
  \label{eq:cutoff}
\end{align}
and
\begin{align*}
  &\left|\frac{1}{4\pi}\int_{\mathbf{S}^1}\int_{d<|s|<R}
    \Rad_\theta f(s)\, D\Hil\Rad_\theta\varphi(s)
    \, ds\, d\theta\right|
  \leq \frac{C_\varphi}{d}
    \sup_{\theta,\,|s|>d} |\Rad_\theta f(s)|,
\end{align*}
where
\[
  C_\varphi = C_\varphi(d)
  = \sup_{|s|>d} s^2 |D\Hil\Rad_\theta\varphi(s)|
  = \sup_{|s|>d}
    \frac{2s^2}{\pi\bigl(|s|+(s^2-1)^{1/2}\bigr)(s^2-1)^{1/2}}.
\]
\end{proposition}

\begin{proof}
Using~\eqref{eq:der-hil-rad-phi} in Properties~\ref{prop:disk}:
\begin{align*}
  &\left|\frac{1}{4\pi}\int_{\mathbf{S}^1}\int_{|s|>d}
    \Rad_\theta f(s)\, D\Hil\Rad_\theta\varphi(s)
    \, ds\, d\theta\right| \\
  &\qquad\leq \frac{C_\varphi}{4\pi}
    \int_{\mathbf{S}^1}\int_{|s|>d}
    \frac{|\Rad_\theta f(s)|}{s^2} \, ds\, d\theta \\
  &\qquad\leq \frac{C_\varphi}{4\pi d^2}
    \int_{\mathbf{S}^1}\int_{|s|>d} |\Rad_\theta f(s)| \, ds\, d\theta.
\end{align*}
The second estimate follows in the same way.
\end{proof}

\subsubsection*{Remarks}

\textit{First}, estimates in terms of $f$ rather than $\Rad_\theta f$ follow
from: $|\Rad_\theta f| \leq \Rad_\theta|f|$; $\int\Rad_\theta f(s)\,ds =
\int f(x)\,dx$; and, for $f$ supported in $|x|<R$,
$|\Rad_\theta f(s)| \leq 2(R^2-s^2)^{1/2}\sup_{d<|x|<R}|f(x)|$ for
$d < |s| < R$.

\medskip

\textit{Second}, using H\"{o}lder's inequality yields intermediate estimates
between the two stated ones; an interpolation argument may be used 
for estimates in $f$.

\medskip

\textit{Third}, consider rescaling.
Let $\varphi_n = n^2\varphi(n\,\cdot)$, which tends to the Dirac delta.
Since $\Rad_\theta\varphi_n(s) = n\Rad_\theta\varphi(ns)$ and the Hilbert
transform is scale-invariant, the estimates in
Proposition~\ref{prop:cutoff} remain unchanged when $\varphi_n$ replaces
$\varphi$, but with $C_\varphi(nd)$ in place of $C_\varphi(d)$.
In the limit, pointwise estimates are obtained with constant $1/\pi$ in
place of $C_\varphi(nd)$, consistent with the pointwise
formula~\eqref{eq:fo} below.

\bigskip

\textit{Fourth}, recall $\Rad_\theta D_i f = \theta_i D\Rad_\theta f$, so
the derivative may be moved from $f$ to $\varphi$ in~\eqref{eq:conv},
assuming $\varphi \in \Schwartz$.
By Proposition~\ref{prop:damping}, the kernel
$\theta_i\Hil D^2\Rad_\theta\varphi(-s)$ then decays as $|s|^{-3}$,
giving an extra factor of $1/d$ in the cut-off error for the mean value
of $D_i f$ at $x=0$:
\[
  \left|\frac{1}{4\pi}\int_{\mathbf{S}^1}\int_{|s|>d}
    \Rad_\theta f(s)\,\theta_i\Hil D^2\Rad_\theta\varphi(-s) \, ds\,
    d\theta\right|
  \leq \frac{C_\varphi}{4\pi d^3}
    \int_{\mathbf{S}^1}\int_{|s|>d} |\Rad_\theta f(s)| \, ds\, d\theta,
\]
where
$C_\varphi = C_\varphi(d) =
\sup_{\theta,|s|>d} |s|^3|\theta_i\Hil D^2\Rad_\theta\varphi(s)|$.
A scaling argument as above leaves the estimate unchanged but for the
substitution $C_\varphi(d) \to C_\varphi(nd)$; in the limit the constant
is $2/\pi$.
Note that if $\varphi$ is even then $\theta_i\Hil D^2\Rad_\theta\varphi(s)$
is odd in both $\theta$ and $s$, giving cancellations; when $f$ is also
even, the entire contribution from $|s|>d$ vanishes.

  We may also use the argument for derivatives of higher orders.
   Recall that ($\varphi\in\Schwartz$): $ \Rad _{\theta} x\cdot\nabla \varphi =
     x\cdot\theta D \Rad _{\theta} \varphi $.
   Now, by induction ($k\geq 0$):
   \[ \Rad _{\theta} (x\cdot\nabla)^k \varphi =
     (x\cdot\theta)^k D^k \Rad _{\theta}  \varphi \, .    
   \]
   Note further that, in a power series expansion of a smooth
   function $h$ around the origin,
   the term of order $k$ is:
   \[ 1/k!\, (x\cdot\nabla)^k h (0) \, .
   \]
We define the cut-off contribution to $f(x)$ in the disk $|x|<d$ from
lines not intersecting that disk, \textit{cf.} \eqref{eq:backproj}:
\begin{equation}\label{eq:fo}
  f_o(x) := \frac{1}{4\pi}\int_{\mathbf{S}^1}\int_{|s|>d}
    \frac{-1}{\pi(x\cdot\theta-s)^2}\, \Rad_\theta f(s) \, ds\, d\theta,
    \qquad |x| < d.
\end{equation}
(The definition of $f_o$ is related to the Fantappi\`{e} transform;
see~\cite{andersson2004}.)

Reasoning as above (alternatively using \eqref{eq:fo} directly), we get ($|x|<d$):
 \begin{eqnarray*} |1/k!\, (x\cdot\nabla)^k f_o (0)| & \leq  &  
     (k+1)/(4\pi^2 d^{k+2}) \int_{\bf S} \int_{|s|>d}  | (x\cdot\theta)^k
                                                           \Rad _{\theta} f (s)|\, \, ds  \, d\theta \\
     \ \ \ \  & \leq  & (k+1) |x|^k/(4\pi^2 d^{k+2}) \int_{\bf S} \int_{|s|>d}
                        |\Rad _{\theta} f (s)|\, \, ds  \, d\theta \, .
   \end{eqnarray*}
   Thus, the power series expansion
   $\sum_{k\geq 0} 1/k!\, (x\cdot\nabla)^k f_o (0)$ converges in the disk $|x|<d$,
   so $f_o(x)$ is an analytic
   function there. Summing the estimates, we get:
\[
  |f_o(x)| \leq \frac{1}{4\pi^2(d-|x|)^2}
    \int_{\mathbf{S}^1}\int_{|s|>d} |\Rad_\theta f(s)| \, ds\, d\theta.
\]
This also follows directly from the definition of $f_o$.
We recapitulate as a proposition.

\begin{proposition}\label{prop:fo}
The function $f_o(x)$ defined in~\eqref{eq:fo}, representing the
contribution to $f(x)$ in $|x|<d$ from lines not intersecting that disk,
is analytic in $|x|<d$, and for $|x|<d$:
\[
  |f_o(x)| \leq \frac{1}{4\pi^2(d-|x|)^2}
    \int_{\mathbf{S}^1}\int_{|s|>d} |\Rad_\theta f(s)| \, ds\, d\theta.
\]
The remaining contribution $f(x) - f_o(x)$ from lines intersecting the
disk is likewise smooth in $|x|<d$.  In particular:
\begin{align*}
  |f_o(0)|
    &\leq \frac{1}{4\pi^2 d^2}
      \int_{\mathbf{S}^1}\int_{|s|>d} |\Rad_\theta f(s)| \, ds\, d\theta, \\
  |f_o(0)|
    &\leq \frac{1}{\pi d}
      \sup_{\theta,\,|s|>d} |\Rad_\theta f(s)|, \\
  |D_i f_o(0)|
    &\leq \frac{1}{2\pi^2 d^3}
      \int_{\mathbf{S}^1}\int_{|s|>d} |\Rad_\theta f(s)| \, ds\, d\theta, \\
  |D_i f_o(0)|
    &\leq \frac{1}{4\pi d^2}
      \sup_{\theta,\,|s|>d} |\Rad_\theta f(s)|.
\end{align*}
\end{proposition}

The estimates in Propositions~\ref{prop:cutoff} and~\ref{prop:fo} do not
account for cancellation in $\Rad_\theta f(s)$; such cancellations would
make the errors smaller.
Writing $f = f_+ - f_-$ with $f_\pm = \max(\pm f, 0)$, the cut-off
error~\eqref{eq:cutoff} is at most the maximum of the corresponding
errors for $f_+$ and $f_-$ separately.
Hence we need only consider $f \geq 0$.

Note that for non-negative $f$ supported in $|x|<R$, the weight function
in~\eqref{eq:fo} has constant sign and decreasing modulus; the error is
monotone in $f$ and maximized when $f(x) = \sup_{d<|x|<R} f(x)$ on the
annulus $d < |x| < R$.
That the $d^{-2}$ decay rate in Proposition~\ref{prop:cutoff} cannot be
improved is seen by placing a unit point mass at distance $r > d$; a
direct calculation shows that the decay rate is sharp. This also holds
for the supremum norm estimate.

\bigskip

We further restrict to $f \geq 0$ supported in $|x|<R$, and consider two
ways to induce cancellation and reduce the cut-off error.

\subsubsection*{Estimated overall mean value}

One possibility to induce cancellation is to subtract Radon data corresponding to an estimated
overall mean value --- computed from the enlarged-region-of-interest data only --- before
reconstruction, then add it back.
Consider the enlarged region $|x|<r<R$.
Integrating $\Rad_\theta f(s)$ over $|s|<r$ gives the integral of $f$
over a cross-section of $|x|<R$ symmetric about the diameter orthogonal
to $\theta$, which always contains $|x|<r$.
Averaging over $\theta$ and dividing by the section area gives an
approximate overall mean of $f$, biased toward $|x|<r$.

An extreme case illustrates the procedure.
Let $f = \mathbf{1}_{|x|<R}$ with data cut off outside $|x|<r<R$.
The approximate mean equals $1$; subtracting the corresponding Radon data
gives $0$, which reconstructs to $0$.
Adding back the mean value recovers $f$ exactly --- yet the cut-off
error~\eqref{eq:cutoff} for this $f$ is maximal.

\subsubsection*{Local means}

A local alternative is to subtract a constant --- the \textit{local mean}
--- at each lattice point before reconstruction, and then add it back. This
procedure will not appear further in our considerations.

At $x = 0$, there is a unique $\mu = \mu_0$ solving the linear equation
\[
  \int_{\mathbf{S}^1}\int_{d<|s|<R}
    \Rad_\theta(f - \mu\,\mathbf{1}_{D_R})(s)\,
    D\Hil\Rad_\theta\varphi(s) \, ds\, d\theta = 0,
\]
where $D_R$ denotes the disk $|x|<R$. Clearly, the choice $\mu = \mu_0$ makes
the cut-off error~\eqref{eq:cutoff} vanish.

Using~\eqref{eq:der-hil-rad-phi}, the solution is:
\begin{align*}
  \mu_0
    &= \int_{\mathbf{S}^1}\int_{d<|s|<R}
       \frac{\Rad_\theta f(s)}%
         {\bigl(|s|+(s^2-1)^{1/2}\bigr)(s^2-1)^{1/2}}
       \, ds\, d\theta \\
    &\quad\Bigg/\,
       \int_{\mathbf{S}^1}\int_{d<|s|<R}
       \frac{\Rad_\theta\mathbf{1}_{D_R}(s)}%
         {\bigl(|s|+(s^2-1)^{1/2}\bigr)(s^2-1)^{1/2}}
       \, ds\, d\theta.
\end{align*}
At a general point $x$, substitute $|x\cdot\theta - s|$ for $|s|$ in the
even denominator kernel, and $d < |x\cdot\theta - s|$ for the domain of
integration, to obtain $\mu_x$.

In principle, the procedure also applies to the pointwise cut-off
error~\eqref{eq:fo}.

\subsubsection*{Remark on dimensions $n > 2$}

From Section~\ref{sec:balls}, cut-off is relevant only in even dimensions $n = 2k$,
since this is precisely when the Hilbert transform is operative in~\eqref{eq:conv-prop}.
We must use $D^{k-1}\Rad_\theta f$ in the reconstruction formula.
Proceeding as in Proposition~\ref{prop:cutoff}, the $d^{-2}$ decay is
replaced by $d^{-(k+1)}$, and the integral of $\Rad_\theta f$ by that of
$D^{k-1}\Rad_\theta f$.
Proposition~\ref{prop:fo} remains valid in even dimensions $n = 2k > 2$,
with modified constants, due the change from $\mathbf S^1$ to $\mathbf S^{n-1}$.

\subsection{Region of interest}\label{sec:roi}

We now turn to the question of how a cut-off to an enlarged region of
interest affects the reconstruction.
We consider the case when the region of interest is a central disk
$|x|<r<R$, with $f$ supported in $|x|<R$.
As argued in Section~\ref{sec:cutoff}, we may restrict to non-negative $f$.

As recalled in the Introduction, the Radon transform may vanish inside a
disk even when $f$ is non-vanishing there.
We discuss how this affects reconstruction in the region of interest;
assume $\varphi \in \Schwartz$.
From the remarks after Proposition~\ref{prop:cutoff}, the kernel
$\Hil D^2\Rad_\theta\varphi(s)$ decays as $|s|^{-3}$, giving an extra
factor of $1/d$ in the cut-off estimate for $D_i f$.
Thus, with $\sup_{r<|x|<R} f(x) = 1$, the contribution from
$\Rad_\theta f$ outside $|x|<r$ to the derivatives $D_i f$ in the region
of interest is bounded roughly by $1/r$ when $R/r = 2$.
Taking cancellations into account --- which occur when $f(x)$ is constant
for $|x|>r$ and $\varphi$ is even, in which case the contribution at $x=0$
vanishes --- we conclude that, when the enlarged region of interest is not
too small:
\begin{itemize}
  \item the Radon data cut off essentially correspond to the addition of
        a constant in the region of interest.
\end{itemize}
(The same observation is made empirically in~\cite{natterer1986}, based on
estimates of $|f_o(x)-f_o(0)|$, see~\eqref{eq:fo}, obtained by numerical
evaluation of integrals.)

Finally, a cut-off makes the
remaining Radon data relatively more emphasized.
For a non-negative function $f$, the contribution to $f(x)$ in $|x|<r$
from lines not intersecting that disk is non-positive,
according to~\eqref{eq:fo}.
The effect is greatest near the cut-off boundary, and it may be mitigated by
weighting the Radon data according to the length of intersection with the
enlarged region. In view of the Support Theorem~\cite{helgason1999},
this procedure is most
effective when the weighted integrals agree with the part intersecting the region.

\section{Numerical experiments}\label{sec:numerics}

We test our methods numerically, as a proof of concept for reconstruction from
data in an enlarged region of interest only.
We compare our region-of-interest reconstruction against filtered back-projections
as implemented in \textsc{Matlab} R2024b, using the \textsc{Matlab} implementation of the
Radon transform, \texttt{radon}, for the common input to all calculations.%
\footnote{The \textsc{Matlab} documentation indicates that this
implementation emulates a beam with width twice the pixel distance and
with linearly decreasing intensity.}

The distance between adjacent Radon lines in the output from the
\textsc{Matlab} function \texttt{radon} equals the
pixel distance.
For the reconstruction density we use the disk of radius $1/\sqrt{\pi}$ (unit
area) in pixel units; the cut-off error estimates of Proposition~\ref{prop:cutoff}
apply, with $C_\varphi$ rescaled to this radius.

Our method is implemented as a straightforward discretization of
formula~\eqref{eq:conv} in dimension $2$, using
properties~\eqref{eq:rad-phi}--\eqref{eq:der-hil-rad-phi} and the mean-value
corrections of Section~3.2.
The singularity of the kernel at $|s|=1$ is handled by assuming that the Radon
transform is piecewise constant on each measurement interval, which yields
a closed-form, singularity-free expression for the computations.

\begin{figure}[htbp]
  \centering
  \begin{subfigure}[t]{0.48\textwidth}
    \centering
    \includegraphics[width=\textwidth]{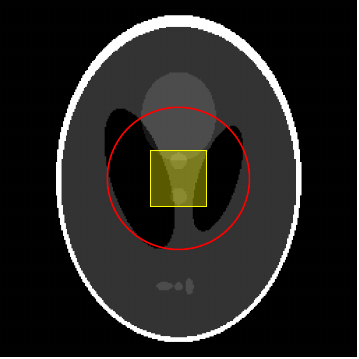}
    \caption{Point phantom}
    \label{fig:phantom-point}
  \end{subfigure}
  \hfill
  \begin{subfigure}[t]{0.48\textwidth}
    \centering
    \includegraphics[width=\textwidth]{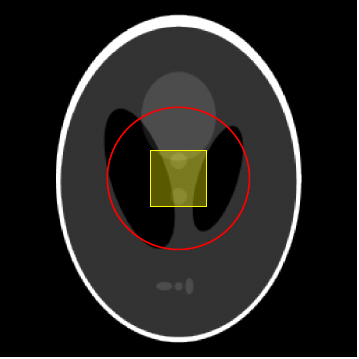}
    \caption{Box-averaged phantom}
    \label{fig:phantom-box}
  \end{subfigure}
  \caption{The $256\times 256$ Modified Shepp--Logan phantom used in the numerical experiments:
    (a) point samples and (b) the box-averaged (anti-aliased) version. The
    semi-transparent yellow square marks the central $40\times 40$ region of
    interest in which quality is assessed; the red circle marks the enlarged
    region of radius $51$ pixels (normalized radius $0.4$), where the Radon data
    are available.}
  \label{fig:phantoms}
\end{figure}

As a test image we use the Modified Shepp--Logan Phantom at $256\times 256$ resolution
(Figure~\ref{fig:phantoms}), available in \textsc{Matlab}, with the \texttt{radon} function applied at
$403$ equally spaced angles over $[0,180)$ --- the standard angular Nyquist rate being
$\pi N/2$. 
These Radon data provide the common input to all reconstructions.

For the filtered back-projection, we use \texttt{iradon} with
\texttt{outputSize}~$= 256$ and nearest-neighbour interpolation
(\texttt{`nearest'}), matching the present absence of interpolation in our method;
all other \texttt{iradon} options are at their default values.

The region of interest, where we assess quality, is the central $40\times 40$
square, that is, $\pm 20$ pixels ($\pm 0.16$ in the normalized coordinates in which
the image spans $[-1,1]$) about the image centre.
The enlarged region, where the Radon data are available, is the central disk $|x|<r$ of
radius $51$ pixels (normalized radius $0.4$) --- only $1.8$ times the half-diagonal
of the region of interest, and containing $28\%$ of the radial samples of the
complete data. For the disk $|x|<R$ of the
estimates above we take $R$ to be the image circle radius equal to $128$ pixels
(normalized radius $1$), of which the enlarged region radius is two fifths. This contains
the Phantom but is not the smallest such disk: the outer ellipse has semi-axes
$(0.69, 0.92)$, so the support is in fact contained in a circle of normalized
radius $0.92$, some $117$ pixels. Either choice is admissible --- the estimates
require only $\operatorname{supp} f \subset \{|x|<R\}$ --- and the procedure is
consistent in each, since the constant $\mu\,\mathbf{1}_{|x|<R}$ is by
construction assigned the measured strip-integrated projections whichever $R$ is
used; we retain the image circle.

We present the results of several reconstructions. Two use the full data and serve
as ceilings:
\begin{itemize}
  \item \textsc{Matlab} filtered back-projection, full data.
  \item Our mean-value reconstruction $f*\varphi$, full data.
\end{itemize}
and, from the truncated data $|s| \leq r$ alone, two pairs, each an
uncorrected reconstruction beside its corrected counterpart:
\begin{itemize}
  \item \textsc{Matlab} filtered back-projection, and the same with the
        estimated overall mean subtracted and restored.
  \item Our $f*\varphi$, and the same with the estimated overall mean.
\end{itemize}

Quality is measured by a linear regression of each reconstruction to a reference
image within the region of interest, minimizing the least-squares residual with
respect to a scale parameter and a level (bias) parameter.
We determine the least-squares solution to
the over-determined system of linear equations $Ax = b$, where $A$ is
$1600 \times 2$ (first column: reconstructed values in the
region of interest;
second column: all ones), $x = (\text{scale}, \text{bias}) \in \mathbf{R}^2$,
and $b$ contains the reference values in the region of interest.
We report the \emph{relative} residual $\|Ax - b\|/\|b\|$ (with $x$ the solution
of $A^tAx=A^tb$, and $\|b\|$ denoting the Euclidean norm of $b$)
together with the fitted scale and bias; normalizing by $\|b\|$
makes the residual comparable across image sizes and region areas. A faithful
reconstruction has scale $\approx 1$ and bias $\approx 0$, so the bias directly
measures the residual level (mean-value) error that the residual norm alone does
not expose.

Because filtered back-projection targets point values whereas our method
reconstructs mean values over a pixel disk, the choice of reference is not neutral. We
therefore report against three references: (i) the point Phantom (pixel values);
(ii) the box-averaged Phantom (each value averaged over its pixel cell), and (iii) the disk-averaged Phantom
(each value averaged over its pixel disk).

\subsubsection*{Results}

Tables~\ref{tab:res-point}, ~\ref{tab:res-box}, and~\ref{tab:res-disk} give the relative residuals,
scales, and biases against the three references. Each groups the two full-data
reconstructions --- which serve as the ceilings for their respective methods ---
above the four computed from enlarged-region-of-interest data only.

\begin{table}[htbp]
  \centering
  \caption{Quality against the \emph{point} Phantom (pixel values). Every
           region-of-interest row uses the truncated data $|s| \leq r$; the rows
           differ only in the reconstruction operator and the correction.
           ``uncorrected'' means no mean is subtracted, ``mean-corrected'' that
           the estimated overall mean of Section~\ref{sec:cutoff} is subtracted
           and restored. The rule separates the two operators.}
  \label{tab:res-point}
  \begin{tabular}{@{}lccc@{}}
    \toprule
    Method & Rel.\ residual & Scale & Bias \\
    \midrule
    \multicolumn{4}{@{}l}{\emph{Full data}} \\
    \quad \textsc{Matlab} FBP              & 0.0882 & 1.0169 & $-0.0019$ \\
    \quad $f * \varphi$                    & 0.0962 & 1.0182 & $-0.0023$ \\
    \addlinespace
    \multicolumn{4}{@{}l}{\emph{Enlarged region-of-interest data only}} \\
    \quad \textsc{Matlab} FBP, uncorrected           & 0.1001 & 1.0294 & $-0.1423$ \\
    \quad \textsc{Matlab} FBP, mean-corrected        & 0.0893 & 1.0119 & $-0.0136$ \\
    \cmidrule{1-4}
    \quad $f * \varphi$, uncorrected                 & 0.1065 & 1.0310 & $-0.1430$ \\
    \quad $f * \varphi$, mean-corrected              & 0.0974 & 1.0131 & $-0.0139$ \\
    \bottomrule
  \end{tabular}
\end{table}

\begin{table}[htbp]
  \centering
  \caption{Quality against the \emph{box-averaged} Phantom; conventions as in
           Table~\ref{tab:res-point}.}
  \label{tab:res-box}
  \begin{tabular}{@{}lccc@{}}
    \toprule
    Method & Rel.\ residual & Scale & Bias \\
    \midrule
    \multicolumn{4}{@{}l}{\emph{Full data}} \\
    \quad \textsc{Matlab} FBP              & 0.0689 & 1.0037 & $\phantom{-}0.0000$ \\
    \quad $f * \varphi$                    & 0.0693 & 1.0064 & $-0.0006$ \\
    \addlinespace
    \multicolumn{4}{@{}l}{\emph{Enlarged region-of-interest data only}} \\
    \quad \textsc{Matlab} FBP, uncorrected           & 0.0834 & 1.0160 & $-0.1385$ \\
    \quad \textsc{Matlab} FBP, mean-corrected        & 0.0702 & 0.9988 & $-0.0116$ \\
    \cmidrule{1-4}
    \quad $f * \varphi$, uncorrected                 & 0.0828 & 1.0191 & $-0.1396$ \\
    \quad $f * \varphi$, mean-corrected              & 0.0707 & 1.0014 & $-0.0120$ \\
    \bottomrule
  \end{tabular}
\end{table}

\begin{table}[htbp]
  \centering
  \caption{Quality against the \emph{disk-averaged} Phantom (each value
           averaged over a disk of radius $1/\sqrt\pi$, the same density as
           $\varphi$); conventions as in Table~\ref{tab:res-point}.}
  \label{tab:res-disk}
  \begin{tabular}{@{}lccc@{}}
    \toprule
    Method & Rel.\ residual & Scale & Bias \\
    \midrule
    \multicolumn{4}{@{}l}{\emph{Full data}} \\
    \quad \textsc{Matlab} FBP              & 0.0646 & 1.0006 & $\phantom{-}0.0004$ \\
    \quad $f * \varphi$                    & 0.0635 & 1.0035 & $-0.0002$ \\
    \addlinespace
    \multicolumn{4}{@{}l}{\emph{Enlarged region-of-interest data only}} \\
    \quad \textsc{Matlab} FBP, uncorrected           & 0.0801 & 1.0128 & $-0.1377$ \\
    \quad \textsc{Matlab} FBP, mean-corrected        & 0.0658 & 0.9957 & $-0.0111$ \\
    \cmidrule{1-4}
    \quad $f * \varphi$, uncorrected                 & 0.0782 & 1.0161 & $-0.1388$ \\
    \quad $f * \varphi$, mean-corrected              & 0.0649 & 0.9985 & $-0.0116$ \\
    \bottomrule
  \end{tabular}
\end{table}

Three observations. \textit{First,} with \emph{full} data our mean-value reconstruction is
comparable to filtered back-projection: it essentially ties it against the
box-averaged Phantom ($0.0693$ vs $0.0689$) and is some nine percent behind
against the point Phantom ($0.0962$ vs $0.0882$), as expected, since
back-projection targets point values whereas $f*\varphi$ is an average over a
disk. The disparity between the two references is itself explained by the choice
of density: at radius $1/\sqrt{\pi}$ the disk has \emph{unit area}, so
$f*\varphi$ and the box-averaged Phantom are averages of $f$ over unit-area
sets --- a disk and a square. Their targets therefore very nearly coincide, and
the remaining difference reflects the shape of the averaging set rather
than any deficiency of the reconstruction; against the point Phantom, by
contrast, the averaging is charged in full.

Against the disk-averaged Phantom --- the reference matched to $f*\varphi$, since a
disk mean is what the method computes --- the ranking reverses: with full data
$f*\varphi$ attains $0.0635$ against back-projection's $0.0646$, and from
region-of-interest data with the correction, $0.0649$ against $0.0658$. The
crossover lies between the box- and disk-averaged references, the box being
near-neutral since it and $\varphi$ both average over unit area.

\textit{Second,} restricting the data to the enlarged region costs comparatively little in
\emph{shape} but a great deal in \emph{level}. For $f*\varphi$ the relative
residual rises from $0.0962$ to $0.1065$ (point) and $0.0693$ to $0.0828$ (box),
between a tenth and a fifth, while the bias degrades from $-0.0023$ to $-0.1430$
and from $-0.0006$ to $-0.1396$ --- by factors of sixty and two hundred.
This is the cut-off error of Section~\ref{sec:cutoff} in numerical
form, and it is, as predicted there, almost entirely a constant: it barely affects
the residual once a level is fitted, but it displaces the absolute values badly.
The cut-off back-projection carries the same level bias, $-0.1423$ and $-0.1385$.

The scale and bias columns separate the two error mechanisms directly. A\-gainst
the matched box-averaged reference the full-data reconstructions have scale
$1.0037$ and $1.0064$ with bias $0.0000$ and $-0.0006$: their error is almost
purely \emph{multiplicative}, a uniform amplitude deficit of a few tenths of a
percent from the residual loss of resolution, with no additive part. The
region-of-interest reconstructions have comparable scales but an additive error
some two orders of magnitude larger. That the data cut-off contributes a constant
while the loss of resolution contributes a factor is precisely the conclusion of
Section~\ref{sec:cutoff}, here visible in the two columns separately.

\textit{Third,} the estimated overall mean reduces the truncation error, and it does so for either
reconstruction operator. Applied to $f*\varphi$ its bias is $-0.0139$ and
$-0.0120$, against the uncorrected $-0.1430$ and $-0.1396$; applied to filtered
back-projection, $-0.0136$ and $-0.0116$ against $-0.1423$ and $-0.1385$. In both
cases this is an order of magnitude smaller than the uncorrected bias, and in both
cases the residual improves as well.

The comparison this invites is each corrected reconstruction against \emph{its
own} ceiling, since the two operators do not target the same thing. From
region-of-interest data alone, $f*\varphi$ attains $0.0974$ against its full-data
$0.0962$ (point) and $0.0707$ against $0.0693$ (box); back-projection attains
$0.0893$ against $0.0882$ and $0.0702$ against $0.0689$. Every corrected
reconstruction is thus within about two percent of what the same operator achieves
with the complete data, on 28 percent of the radial data --- the
truncation penalty is not merely reduced but largely removed. All fitted scales lie within two percent of unity.

That the correction works equally for both operators is to be expected: it
addresses the constant left by truncating the \emph{data}, and any reconstruction
that returns a constant from the Radon data of a constant admits it. The
mean-value formulation is what allows that constancy to be established
(Propositions~\ref{prop:cutoff} and~\ref{prop:fo}); the remedy it suggests is not
confined to that formulation. We note also that the near-parity between the two
operators here is specific to properly sampled data: with angular sampling well
below the above Nyquist rate, the non-singular Radon mean-value kernel we use is
markedly less sensitive to aliasing compared to the ramp filter, and $f*\varphi$ in
Table 4 gives 12 percent against ROI-corrected FBP and 10 percent against full-data FBP, both at half of Nyquist.

We present results from this angular sampling aspect next.

\subsubsection*{Angular sampling}

The near-parity of the two operators above is specific to properly sampled data.
The mean-value Radon density we use is non-singular
in contrast to the ramp filter of
back-projection. As the angular sampling is reduced the aliasing this
introduces penalises back-projection more heavily, and the ordering reverses.
Table~\ref{tab:views} follows the comparison down from the standard Nyquist rate
$\pi N/2$ of the preceding tables, against the box-averaged Phantom.

\begin{table}[htbp]
  \centering
  \caption{Effect of reducing the angular sampling. Relative residuals against
           the box-averaged Phantom. The last two columns are both computed from
           region-of-interest data with the estimated overall mean, and so differ
           only in the reconstruction operator; the first is back-projection given
           the \emph{complete} data, and serves as the reference.}
  \label{tab:views}
  \begin{tabular}{@{}rlccc@{}}
    \toprule
          & Fraction   & Full data & \multicolumn{2}{c}{Region of interest, mean-corrected} \\
    \cmidrule(lr){3-3} \cmidrule(l){4-5}
    Views & of Nyquist & \textsc{Matlab} FBP & \textsc{Matlab} FBP & $f*\varphi$ \\
    \midrule
    403 & $1$   & 0.0689 & 0.0702 & 0.0707 \\
    302 & $3/4$ & 0.0683 & 0.0696 & 0.0710 \\
    269 & $2/3$ & 0.0718 & 0.0734 & 0.0718 \\
    202 & $1/2$ & 0.0899 & 0.0917 & 0.0809 \\
    \bottomrule
  \end{tabular}
\end{table}

Two tentative readings may be made. Comparing the last two columns, which differ only in
the operator applied to the same truncated data, back-projection is ahead at and
just below the Nyquist rate and $f*\varphi$ from two thirds downwards. Comparing
the last column with the first, the mean-value reconstruction from
region-of-interest data alone equals back-projection given \emph{all} of the data
at two thirds of the Nyquist rate ($0.0718$ against $0.0718$), and is ten percent
better at one half ($0.0809$ against $0.0899$). The transition is gradual rather than sharp --- these figures
carry one or two percent of variation with the particular set of angles chosen,
so the crossing should be read as lying somewhere between two thirds and three
quarters of the Nyquist rate --- and it occurs later, nearer one half, when
measured against the point Phantom, which favours back-projection throughout. Note that the FBP-results are not
consistently increasing as the number of views decreases. These deviations seem specific to the 
\textsc{Matlab} FBP-implementation \texttt{iradon}.

A second effect works in the same direction. The penalty for restricting the data
to the enlarged region diminishes as the sampling is reduced: with the correction,
 the truncation error
seems to become small beside the aliasing. Reducing the number of views and
narrowing the measured region seem therefore not competing economies; each makes
the other cheaper.


\bibliographystyle{unsrt}
\bibliography{bibliography}

\end{document}